\documentclass[showpacs,aps,floatfix,twocolumn,preprintnumbers,nofootinbib]{revtex4}

\usepackage{amssymb}
\usepackage{bm}
\usepackage{graphicx}

\begin{document}

\title{Universes out of almost empty space}
\author{Stefano Ansoldi}
\email{ansoldi@trieste.infn.it}
\homepage{http://www-dft.ts.infn.it/~ansoldi}
\affiliation{International Center for Relativistic Astrophysics (ICRA), Pescara, Italy\\
Department of Physics, Kyoto University, Kyoto, Japan}
\altaffiliation{(\emph{Mailing address\/}) Dipartimento di Matematica e Informatica - Universit\`{a} degli Studi di Udine, via delle
Scienze 206, I-33100 Udine (UD), Italy}
\altaffiliation{INFN, Sezione di Trieste, Trieste, Italy}
\author{Eduardo I. Guendelman}
\email{guendel@bgumail.ac.il}
\affiliation{Department of Physics, Ben Gurion Univeristy, Beer Sheva, Israel}

\begin{abstract}
Baby universes (inflationary or non--inflationary)
are regions of spacetime that disconnect from the original ambient
spacetime, which we take to be asymptotically flat spacetime.
A particular kind of baby universe solution, involving string--like
matter, is studied to show that it can be formed by ``investing'' an arbitrarily
small amount of energy, i.e. it can appear from an almost flat space
at the classical level. Since this possibility has not yet been clearly
recognized in the literature, we then discuss in detail its properties, relevance
and possible generalizations.
\end{abstract}

\preprint{KUNS-2095}


\maketitle

In inflationary cosmology \cite{bib:PhReD1981..23...347G,bib:PhLeB1982.108...389L,bib:PhReL1982..48..1220S}
the consideration of the dynamics of an isolated vacuum bubble
\cite{bib:PhReD1977..15..2929C,bib:PhReD1977..16..1248C,bib:PhReD1977..16..1248C,bib:PhReD1977..16..1762C,
bib:PhReD1980..21..3305L,bib:PrThP1981..65..1443M,bib:PhReD1987..35..1747G,bib:PhReD1987..36..2919T},
for instance a de~Sitter bubble \cite{1917PrKoNeA19..1217..D,1917PrKoNeA20..229...D}
in an external Schwarzschild \cite{bib:Schwarzschild1} space,
leads to the concept of \emph{baby universe}: for a big enough
bubble of false vacuum, the expansion to arbitrarily large values of the radius
is inevitable despite the fact that the pressure difference (negative inside, zero outside)
and the existence of a non-exotic (i.e. positive) surface tension prevent the bubble from expanding \emph{into}
the external space. The presence of wormholes makes possible this feature
\cite{bib:PhReD1981..23...347G,bib:Tomimatsu:1990aa}, which is, otherwise,
counterintuitive: due to the peculiar structure of the Kruskal extension \cite{bib:PhRev1960.119..1743K},
the bubble can, in fact, expand \emph{by making its own space}, provided it is located in the wormhole
region of Schwarzschild space. Then, the mechanical force that accelerates the wall,
which is pushing from outside to inside, acts in the direction of increasing radius and supports the
bubble expansion! This solves the apparent paradox showing that
consistent solutions that describe inflationary bubbles expanding to infinity
in asymptotically flat spacetimes are baby universes.
There are two features common to most of the baby universe models present in the literature:
the first one is the fact that they have been considered, likely for historical reasons, in the
context of the inflationary scenario; the second one is that a generic feature of the existing models is
the presence of a critical value $M _{\mathrm{cr}}$ for the total mass--energy in the
asymptotically flat region, below which the bubble cannot expand to infinity \cite{bib:PhReD1987..35..1747G}.
In this letter we would like to show that
baby universes, intended in the broader sense of regions of spacetime
disconnecting from the ambient space, can be considered under more general
settings, i.e. we will not restrict ourself to the framework of inflationary cosmology.
We will also show that it is possible to develop models in which the value of the critical mass
can become arbitrarily small and discuss possible implications and consequences.
\begin{figure*}[!tbh]
\begin{center}
\includegraphics[width=17.9cm]{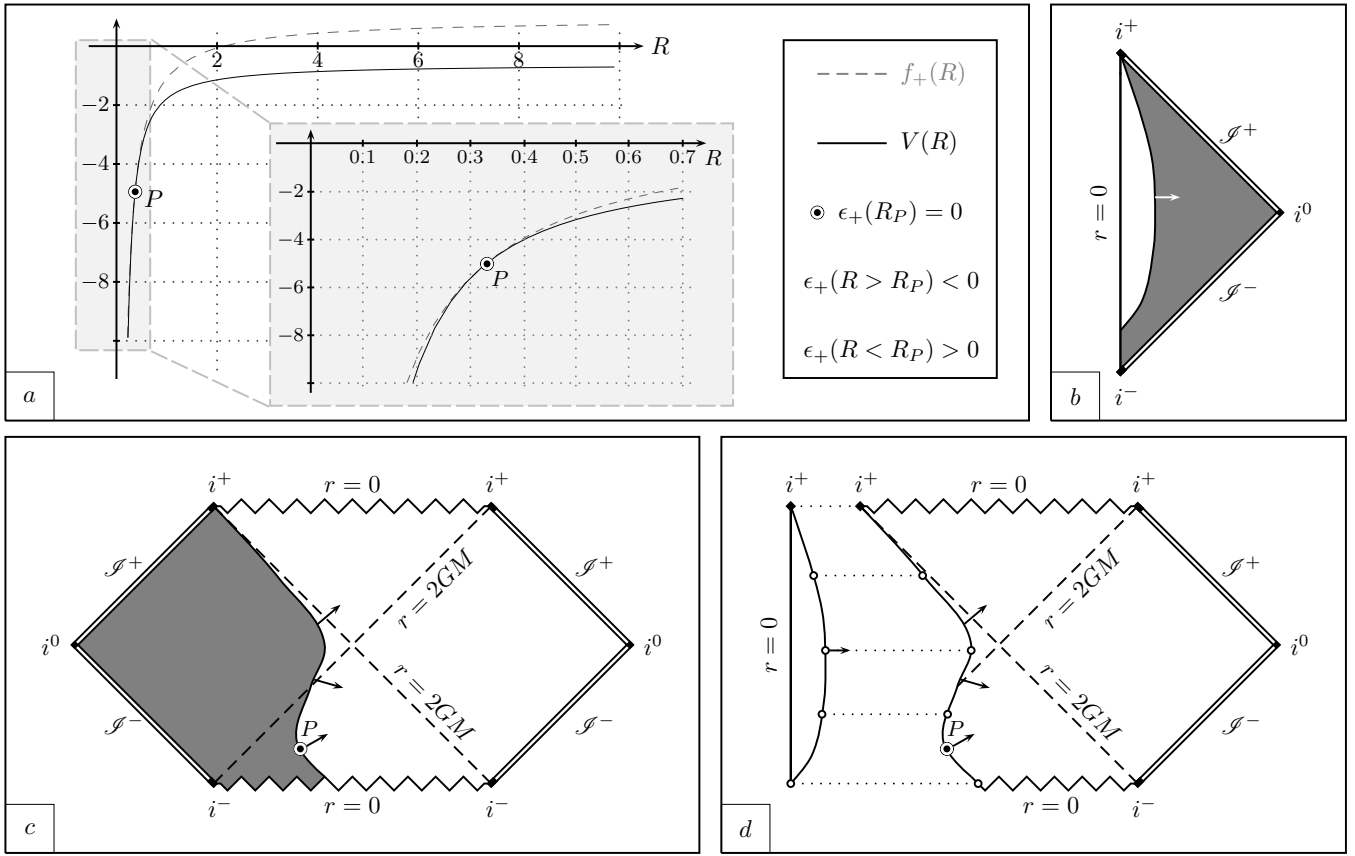}
\caption{\label{fig:001}Construction of the global spacetime structure with the classical
formation of a baby universe for an unbounded solution of the Minkowskii--Schwarzschild
junction by a sphere of strings (boundaries at infinity are labelled according to the conventions
in \cite{bib:CaUPr1973...1...391E}).
In panel ``a'' we plot the effective potential (solid curve)
together with the horizon curve $f _{+} (R) = 1 - 2 G M / R$ (dashed gray curve). The
effective classical dynamics if given by the motion of the representative point along
the $R$ axis, which corresponds to a motion with zero energy in the potential $V (R)$
(cf. eq.~({\protect\ref{eq:effequ}})). Correspondingly the shell starts from zero radius,
expanding toward infinity. In the two spacetimes this motion
corresponds to the solid curves in panels ``b'' and ``c''. Always from panel ``a'', it is transparent
that the shell will cross the white hole horizon, since it will expand up to arbitrarily large
values of the radius $R$. To determine if the part of spacetime participating in the junction will
be the one on the left or on the right of the shell trajectory, we have to look at the normal
(also shown in the picture). In Minkowskii spacetime
(see panel ``b'') the sign $\epsilon _{-}$ is always positive, i.e. the normal always points in the
direction of increasing radii. Moreover our convention is that the normal is going from
${\mathcal{M}} _{-}$ to ${\mathcal{M}} _{+}$ so that the \emph{un}shaded region is, in fact,
${\mathcal{M}} _{-}$. An analogous procedure has to be performed in Schwarzschild spacetime
(panel ``c''). Again ${\mathcal{M}} _{-}$ is the \emph{un}shaded region.
In this case, however, a new feature appears, since the sign $\epsilon _{-}$ changes at $P$,
and the normal that before $P$ was pointing in the direction of decreasing radii, after $P$
will point in the direction of increasing radii. This change between the relative orientation
of the normal and the increasing radii direction happens when the horizon curve $f (R)$ is
tangent to the potential curve $V (R)$ (see the zoomed snapshot in panel ``a''). All these
properties give rise to the global spacetime structure in panel ``d'': the fact that the
asymptotically flat part of Schwarzschild spacetime includes the central point of the diagram,
which in turn is due to the fact that outside the horizon $\epsilon _{+} = -1$, signals that
a baby universe is formed in the process.}
\end{center}
\end{figure*}

Following a consolidate approach we will model the vacuum bubble and the
baby universe formation process using a \emph{thin relativistic shell}
separating two spherically symmetric domains of
spacetime; this description has the advantage of a well known classical formulation
for the spacetime dynamics in terms of Israel junction conditions
\cite{bib:NuCim1966.B44.....1I,1967NuCiB..48..463...I,bib:PhReD1991..43..1129I},
which is free from ambiguities and represents a solid toehold to also
tackle the study of the quantum dynamics with less (but still not
completely solved \cite{Ansoldi:2007xu}) technical problems. This seminal idea has been
proposed long ago by various authors
\cite{bib:PhReD1987..35..1747G,bib:PhReD1987..36..2919T,bib:NuPhy1990B339...417G,%
bib:PhReD1990..41..2638P};
crucial parameters of the model are, apart from the total mass--energy
of the asymptotically flat region, the false vacuum energy
density (when present) and the shell surface tension. Let us then consider
two spacetime domains ${\mathcal{M}} _{\pm}$ of two
$(3+1)$-dimensional spacetimes, separated by an infinitesimally
thin layer of matter $\Sigma$, a \emph{shell}. We will also assume
spherical symmetry: this simplifies the algebra, and is a
non-restrictive assumption which almost always appears in the
literature. Moreover, as a concrete case we will choose the one in
which ${\mathcal{M}} _{-}$ is a part of Minkowskii spacetime and
${\mathcal{M}} _{+}$ is a part of Schwarzschild spacetime. Then,
the equations of motion for the shell, i.e. Israel junction
conditions, reduce to the single equation
\begin{equation}
    \epsilon _{-} \sqrt{\dot{R} ^{2} + 1} - \epsilon _{+} \sqrt{\dot{R} ^{2} + 1 - 2 G M / R} = G m (R) / R ,
\label{eq:juncon}
\end{equation}
where $G$ is the gravitational constant and
the only remaining degree of freedom is $R (\tau)$, the
radius of the spherical shell expressed as a function of the
proper time $\tau$ of an observer co-moving with the shell
(an overdot indicates a derivative with respect to $\tau$).
$\epsilon _{\pm} = \mathrm{sgn} (n ^{\mu} \partial _{\mu} r
)\rceil _{\mathcal{M} _{\pm}}$ determine if
the radial coordinate $r$ is increasing ($\epsilon _{\pm} = +1$) or
decreasing ($\epsilon _{\pm} = -1$) along the normal direction,
defined by $n ^{\mu} \rceil _{{\mathcal{M}} _{\pm}}$ in
${\mathcal{M}} _{\pm}$, respectively (our convention is
that the normal is pointing from ${\mathcal{M}} _{-}$ to
${\mathcal{M}} _{+}$). The function $m(R)$ is related to the energy--matter
content of the shell, and is what remains of the shell stress--energy
tensor in spherical symmetry after
relating the pressure $p$ and the surface energy density $\rho$ \emph{via} an
equation of state. Let us discuss this point in more detail, since
our choice will be slightly unusual compared to the existing
literature. We will, in fact, use $p = - \rho / 2 $, $p$ being the
uniform pressure and $\rho$ the uniform energy density on
$\Sigma$. A string gas in $n$ spatial dimensions satisfies
$p = - \rho / n $, therefore the two dimensional shell $\Sigma$
we are dealing with is a \emph{sphere of strings}.
This, gives $\rho = \rho_0 / R$, where $\rho_0$ is a
constant, and then, $m (R) = c R$, with $c=4\pi \rho_0$.
After making the above choice, we can then solve (\ref{eq:juncon}); this is
equivalent to the solution of the equivalent effective classical problem
\cite{bib:PhReD1987..35..1747G,bib:PhReD1989..40..2511S}
\begin{equation}
    \dot{R} ^{2} + V (R) = 0, \quad
    V (R) = 1 - \frac{1}{4 c ^{2}} \left( \frac{2 M}{R} + G c ^{2} \right) ^{2} ,
\label{eq:effequ}
\end{equation}
with the signs determined as $\epsilon _{-} = + 1$ and $\epsilon _{+} = \mathrm{sgn} ( 2 M / R - G c ^{2} )$.
Moreover the potential satisfies
\[
    \lim _{R \to 0 ^{+}} V (R) =- \infty , \:
    \lim _{R \to \infty} V (R) = 1 - \frac{G ^{2} c ^{2}}{4}, \:
    \displaystyle \frac{d V (R)}{d R} > 0 .
\]
This shows that i) we can have unbounded trajectories
only if $c \geq 2 / G$ and ii) \emph{this is independent from the choice of} $M > 0$. Moreover, the result for
$\epsilon _{+}$, shows that certainly iii) on all the unbounded trajectories $\epsilon _{+}$ changes
sign (being positive for small enough $R$ and negative for large enough $R$; the general property that
this change of sign must happen behind an horizon, is also obviously satisfied since $c > 2 / G$, so that at
$R = 2 G M$ the sign $\epsilon _{+}$ is already negative although for small enough $R$ it is positive).
Then, the global spacetime structure associated with the above properties of \emph{all} the unbounded
solutions (please, see figure \ref{fig:001}), clearly shows that they realize the formation of
a baby universe. This happens for a large enough density of strings and \emph{for any} positive value of $M$.
We thus have a first example of a baby universe that can be classically formed out of almost
empty space.

It is interesting to quickly consider what would have happened if we would have
considered the shell formed by matter of a different kind, with an equation of state resulting
in a function $m(R)$ (see equation~(\ref{eq:juncon})) of the following form:
$m (R) = \alpha R ^{\lambda}$, $\alpha > 0$ and $\lambda$ arbitrary. Then, for $\lambda > 1$
classical production of baby universes can take place out of almost empty space only in
the limit in which $\alpha$ becomes very large (this could provide a specific model of
a very high ultraviolet excitation, which nevertheless has a very low overall mass).
Otherwise a potential barrier appears and
quantum effects are required to let a \emph{bounded} solution tunnel into a \emph{bounce},
expanding one, that describes a baby universe. If $\lambda < 1$, instead,
no baby universe solutions can be obtained, \emph{either} because the potential does not admit solutions
that expand to infinity, \emph{or} because these solutions do not contain a baby universe. In this sense,
the case that we have studied, $\lambda = 1$, is the lowest $\lambda$ example
of baby universe formation out of almost empty space: moreover, it is the only one in which quantum effects
are not required, at least within the class of models considered above. For $\lambda = 1$,
in particular, we can interpret the matter content of the model as a gas of strings
being located on the shell (a realization of what is called a dynamical surface tension). This
dramatically affects the critical mass threshold for the production of baby universes which
is present in other models (for instance, the $\lambda > 1$ ones). In fact, as the
strength of this string shell, parametrized by the constant $c$, is increased
above $2 / G$, the critical mass for baby universe production becomes arbitrarily
small. This means that even a minimal (in terms of energy) quantum
fluctuation could trigger the creation of a baby universe.
The condition $c > 2 / G$ means that the string shells
have $\rho _{0} > 1/(2 \pi G)$, i.e., they are transplanckian. This is
in agreement with arguments that support the notion that at transplanckian scales
baby universe production is unsuppressed \cite{bib:2007GraResFouEdu}.
Although there are already models in which all bubbles starting from zero
radius can grow arbitrarily large\footnote{They introduce a multiplet of scalar fields
having the configuration of a global monopole of big enough
strength, called ``hedgehog'' \cite{bib:PhReD1991..44..3152R}
(the subject of inflation assisted by topological defects was also studied in
\cite{bib:PhLeB1994.327...208L} and \cite{bib:PhReL1994..44..3137V}); it has, moreover, been
shown that this effect also holds in the gauged case
\cite{bib:PhLeB1994.327...208L,bib:PhReL1994..44..3137V,bib:PhReD1991..44..3152R}.
Some more elaborate processes by which supercritical magnetic monopoles can be obtained \cite{bib:PhReD1999..59043513V}
and the problem of initial singularity \cite{bib:PhReD2006..74024026V} have also been studied.},
the question of what is the critical mass to make a universe out
of an asymptotically flat space has
\emph{not been addressed}.

Since baby universes have often been considered as
possible solutions for various interesting problems (see, for instance,
the early \cite{bib:Nielsen:1988kf}),
but the possibility of creating baby universes out of almost empty space
has not been appreciated in the literature so far, we would now like to discuss some
possible consequences of this new result.
These considerations might also partly address the question about the observability/detectability of
baby universes: although in our purely classical example the physics beyond the wormhole
is completely disconnected from the one in the \emph{parent} universe (so that a \emph{direct}
evidence of unsuppressed baby universe production might be difficult to achieve)
an \emph{indirect} one could be acquired \emph{via} secondary effects.

The first of them is that unsuppressed baby universe creation could act as
an effective regulator for ultraviolet divergences in quantum gravity theories: let us consider
a model in which strings and their excitations represent matter quantum states. If baby
universe production is unsuppressed (or, even enhanced, as it is at higher energy density) super-heavy
states, present in the spectrum of the theory or arising in an interaction picture, could be
secluded from our universe and realized as baby universe causally disconnected from our one. A
simplified, lowest order, model of these process could be the merging of two string balls having both,
for instance, subcritical amount of string content, into a new supercritical one.
This could significantly soften ultraviolet problems in quantum gravity. Similar ideas
could be important also when considering the ground state of the theory, since states that
can be produced at no (or \emph{very small}) energy cost can condensate, heavily affecting
the structure of the vacuum and providing new possibilities to address problems like those
related to the value of the cosmological constant (see, for instance, the recent
\cite{bib:Rodrigo:2007zv} for a discussion of the possible role of Minkowskii wormholes).

Another effect that might be important to consider if unsuppressed baby universe
creation would take place, is in connection with the unitarity of the theory and the
problem of information loss \cite{bib:Hawking:1987mz}
(this seems a situation that cannot be neglected when some states of the
theory could be effectively confined inside a newly formed and causally
disconnected baby universe). These ideas have already been considered in the standard
settings, but their consequences would be certainly enhanced if the creation could take
place at low or no energy cost, as we have discussed in this paper.

It is also suggestive to notice that, in a not too far future, it might also be possible to
devise suitable laboratory tests to have a more direct experimental contact with some
of the effects that we summarized above: as a matter of fact, it has already been proposed that
analogue models could be used to also identify in terms of condensed matter degrees of freedom
\cite{bib:Jacobson:2000gs} the relevant signature of baby universe creation; clearly, unsuppressed
baby universe creation would then have a peculiar one; an understanding of this signature
could be decisive, especially nowadays that smaller and smaller length scales are becoming
available to observation; it might, in fact, be able to provide us with more detailed information
about how the universe was in fact created, and if it really disconnected from a pre--existing universe,
as unsuppressed baby universe creation might make more likely.

In view of the above discussed possibilities, it would be interesting to further study and
extend the simple model presented in this paper, by providing a general classification
of models and conditions that allow universe creation at a very small energy cost. In this context
we would like to study the small tunnelling effects that are necessary when
potential barriers for creating a universe are arbitrary low but non-zero,
as well as to show that also using a hedgehog a baby universe can be
consistently produced at small energy cost. Moreover, it would be very interesting to solve
a similar problem with just a string loop, and see whether this can give
rise to a baby universe without the presence of a threshold, i.e. a critical mass:
such a problem is of high (to say the least) mathematical complexity. The above ideas
will be considered in a future publication, where we will present a more general study
of baby universes for which there is no critical mass barrier \cite{bib:InPrep}.

\acknowledgements

EIG, would like to thank the university of Udine for
hospitality and Idan Shilon for stimulating conversations about the subject
of this letter. The work of S. Ansoldi is partly supported by a JSPS
invitation Fellowship.

\end{document}